\documentclass[conference]{IEEEtran}
\usepackage{cite}
\usepackage{amsmath,amssymb,amsfonts}
\usepackage{algorithmic}
\usepackage{graphicx}
\usepackage{textcomp}
\usepackage{xcolor}
\usepackage{balance}
\usepackage{cite}
\usepackage{ragged2e}
\usepackage{multirow}

\begin{document}

\title{Tunneling-Augmented Simulated Annealing for Short-Block LDPC Code Construction}

\author{
\IEEEauthorblockN{Atharv Kanchi}
\IEEEauthorblockA{
Illinois Mathematics and Science Academy \\
United States \\
akanchi@imsa.edu
}
}

\maketitle

\begin{abstract}
Designing high-performance error-correcting codes at short blocklengths is critical for low-latency communication systems, where decoding is governed by finite-length and graph-structural effects rather than asymptotic properties. This paper presents a global discrete optimization framework for constructing short-block linear codes by directly optimizing parity-check matrices. Code design is formulated as a constrained binary optimization problem with penalties for short cycles, trapping-set-correlated substructures, and degree violations. We employ a hybrid strategy combining tunneling-augmented simulated annealing (TASA) with classical local refinement to explore the resulting non-convex space. Experiments at blocklengths 64-128 over the AWGN channel show 0.1-1.3 dB SNR gains over random LDPC codes (average 0.45 dB) and performance within 0.6 dB of Progressive Edge Growth. In constrained regimes, the method enables design tradeoffs unavailable to greedy approaches. However, structural improvements do not always yield decoding gains: eliminating 1906 trapping set patterns yields only +0.08 dB improvement. These results position annealing-based global optimization as a complementary tool for application-specific code design under multi-objective constraints.

\end{abstract}


\section{Introduction}

Error-correcting codes are a foundational component of modern digital communication systems, enabling reliable transmission over noisy channels. While many classical and capacity-approaching constructions perform well at large blocklengths, their performance often degrades significantly in the short-blocklength regime. This limitation is increasingly relevant as emerging applications such as ultra-reliable low-latency communications, satellite IoT, and industrial control systems impose strict constraints on latency, decoding complexity, and blocklength.

At short blocklengths, asymptotic guarantees no longer accurately predict performance. Instead, decoding behavior is dominated by finite-length effects and the local structure of the code's Tanner graph. In particular, short cycles, stopping sets, and trapping sets can severely degrade the performance of iterative decoders, even when minimum distance properties are acceptable. As a result, code design in this regime requires explicit control over graph-theoretic structure rather than reliance on asymptotic optimality.

Low-Density Parity-Check (LDPC) codes are attractive candidates for short-block applications due to their flexible structure and efficient iterative decoding. However, constructing high-performance short LDPC codes remains challenging. Random constructions often yield poor structural properties, while greedy heuristics such as Progressive Edge Growth (PEG) optimize for specific metrics, such as girth, but struggle to simultaneously balance multiple competing design objectives. Moreover, these methods are difficult to adapt to additional constraints such as forbidden subgraph patterns, prescribed degree distributions, or block-structured parity-check matrices.

Because short-block code design involves multiple competing structural objectives, including girth, cycle multiplicity, and degree constraints, greedy constructions optimized for a single metric may produce suboptimal overall decoding performance. This motivates the use of global optimization methods that can balance several structural penalties simultaneously.

From a computational perspective, LDPC code construction can be viewed as a discrete combinatorial optimization problem: given a target blocklength and rate, one seeks a parity-check matrix that optimizes decoding-relevant structural metrics under practical constraints. This problem is inherently high-dimensional, non-convex, and multi-objective, with a rugged optimization landscape dominated by local minima. These characteristics limit the effectiveness of purely local or greedy construction techniques.

In this work, we investigate a global optimization framework for short-block LDPC code construction based on tunneling-augmented simulated annealing (TASA). The method extends classical simulated annealing~\cite{kirkpatrick1983sa} with a decaying tunneling-inspired perturbation term, motivated by the quantum annealing literature~\cite{kadowaki1998qa,martonak2002qa_pimc}, to better escape local minima in the rugged LDPC design landscape. The algorithm is implemented entirely on classical hardware. Combined with classical local refinement, TASA provides an efficient mechanism for exploring the high-dimensional discrete design space.

The primary contributions of this paper are as follows:
\begin{enumerate}
    \item We formulate short-block LDPC code construction as a constrained discrete optimization problem over parity-check matrices, incorporating penalties for short cycles, stopping sets, and degree violations.
    \item We introduce a hybrid optimization framework that combines global exploration via tunneling-augmented simulated annealing with classical local refinement to escape poor local minima.
    \item We empirically evaluate the proposed method at blocklengths between 64 and 128 bits over the AWGN channel, demonstrating 0.1--1.3 dB SNR gains over random LDPC construction at BLER $= 10^{-2}$ and competitive performance (within 0.6 dB) relative to mature Progressive Edge Growth heuristics.
    \item We analyze performance under several constrained design regimes, including forbidden subgraph constraints, block-structured codes, and irregular degree profiles, demonstrating when global optimization enables design tradeoffs unavailable to greedy methods and when structural improvements do not translate to decoding gains.
\end{enumerate}

The remainder of the paper is organized as follows. Section~II reviews related work. Section~III formulates the code construction problem. Section~IV describes the proposed optimization method. Section~V presents the experimental setup. Section~VI reports results, and Section~VII discusses implications and limitations.

\section{Related Work}

\subsection{Classical LDPC Code Construction}

Low-Density Parity-Check (LDPC) codes were introduced by Gallager~\cite{gallager1962ldpc} and later rediscovered by MacKay~\cite{mackay1999good}, demonstrating near-capacity performance under iterative decoding. At asymptotic blocklengths, LDPC codes approach Shannon capacity~\cite{richardson2008modern}, but their performance at short blocklengths depends critically on Tanner graph structure.

For short-block LDPC construction, the Progressive Edge Growth (PEG) algorithm~\cite{hu2005peg} has become a widely adopted greedy heuristic. PEG constructs codes by sequentially adding edges to maximize local girth, producing graphs with favorable cycle distributions. Extensions such as Decoder-Optimized PEG~\cite{li2007dopeg} and Improved PEG~\cite{he2007improvedpeg} refine this approach by incorporating decoding-specific metrics. However, PEG's greedy nature limits its ability to explore global structural tradeoffs or incorporate multiple simultaneous constraints~\cite{ramamoorthy2004shortblock}.

Recent work has explored systematic construction methods for short LDPC codes~\cite{shortblockldpc2018}, including combinatorial designs and structured matrix approaches, though these methods typically target specific algebraic properties rather than direct performance optimization.

\subsection{Graph-Theoretic Performance Metrics}

The performance of iteratively decoded LDPC codes is strongly influenced by graph-theoretic structures. Di et al.~\cite{di2002finite} analyzed stopping sets on the binary erasure channel, demonstrating how small stopping sets limit finite-length performance. Dolecek et al.~\cite{dolecek2010absorbing} characterized absorbing sets and their role in error floor phenomena. Richardson~\cite{richardson2003errorfloor} identified trapping sets as a primary cause of decoding failures at moderate-to-high SNR, motivating explicit avoidance of harmful substructures during code design. Wang et al.~\cite{wang2021desiredgirth} proposed methods for constructing codes with prescribed girth, though computational complexity limits applicability to very short blocks.

These results establish that short-block LDPC performance requires optimization of multiple competing graph metrics beyond minimum distance alone.

\subsection{Optimization-Based Code Design}

Several prior works have applied optimization techniques to LDPC construction. Venkiah et al.~\cite{venkiah2007ip} formulated short-block code design as an integer programming problem, though scalability remains limited. Campello et al.~\cite{campello2007de} employed differential evolution for irregular LDPC design, demonstrating that metaheuristic search can discover codes competitive with algebraic constructions. Yang et al.~\cite{yang2004moderate} proposed a systematic methodology combining algebraic structure with iterative improvement.

However, these approaches have not explored annealing frameworks augmented with tunneling-inspired perturbations. TASA provides a principled alternative by combining the thermal exploration of simulated annealing~\cite{kirkpatrick1983sa} with a decaying tunneling term motivated by quantum annealing~\cite{kadowaki1998qa}, offering more structured escape from local minima than purely random restarts.

\subsection{Physics-Inspired Optimization}

Quantum annealing, introduced by Kadowaki and Nishimori~\cite{kadowaki1998qa}, exploits quantum tunneling to solve combinatorial optimization problems. Simulated quantum annealing~\cite{martonak2002qa_pimc} implements these principles classically via path-integral Monte Carlo. Farhi et al.~\cite{farhi2014qaoa} developed the Quantum Approximate Optimization Algorithm (QAOA) for gate-based quantum computers. Lucas~\cite{lucas2014ising} demonstrated that many NP-hard problems admit natural Ising/QUBO formulations suitable for annealing.

Recent work has begun applying quantum-inspired methods to communication problems~\cite{kasi2020qbp,luisier2025qu4fec}, though to our knowledge, no prior work has applied these techniques specifically to short-block LDPC code construction as a constrained discrete optimization problem over parity-check matrices.

\subsection{Application Context}

The need for high-performance short-block codes is driven by emerging ultra-reliable low-latency communication (URLLC) requirements in 5G systems~\cite{popovski2018communication_slicing}. Durisi et al.~\cite{durisi2016shortpackets} characterized fundamental limits of short-packet communication, demonstrating that finite-blocklength performance depends critically on code structure rather than asymptotic properties. Satellite IoT applications~\cite{desanctis2016satellite_iort} similarly impose strict latency and blocklength constraints, motivating continued research in short-block code design.

\section{Problem Formulation}

\subsection{Code Representation and Constraints}

We represent a binary linear block code of length $n$ and dimension $k$ by its parity-check matrix $\mathbf{H} \in \{0,1\}^{m \times n}$, where $m = n - k$. The code $\mathcal{C}$ is defined as:
\begin{equation}
\mathcal{C} = \{\mathbf{c} \in \{0,1\}^n : \mathbf{H}\mathbf{c}^T = \mathbf{0} \pmod{2}\}
\end{equation}

For LDPC codes, $\mathbf{H}$ is sparse with row weight $w_r$ and column weight $w_c$ satisfying $w_r, w_c \ll n$. The Tanner graph representation associates variable nodes with code bits and check nodes with parity constraints, with edges corresponding to nonzero entries in $\mathbf{H}$.

\begin{figure}[t]
\centering
\includegraphics[width=0.48\textwidth]{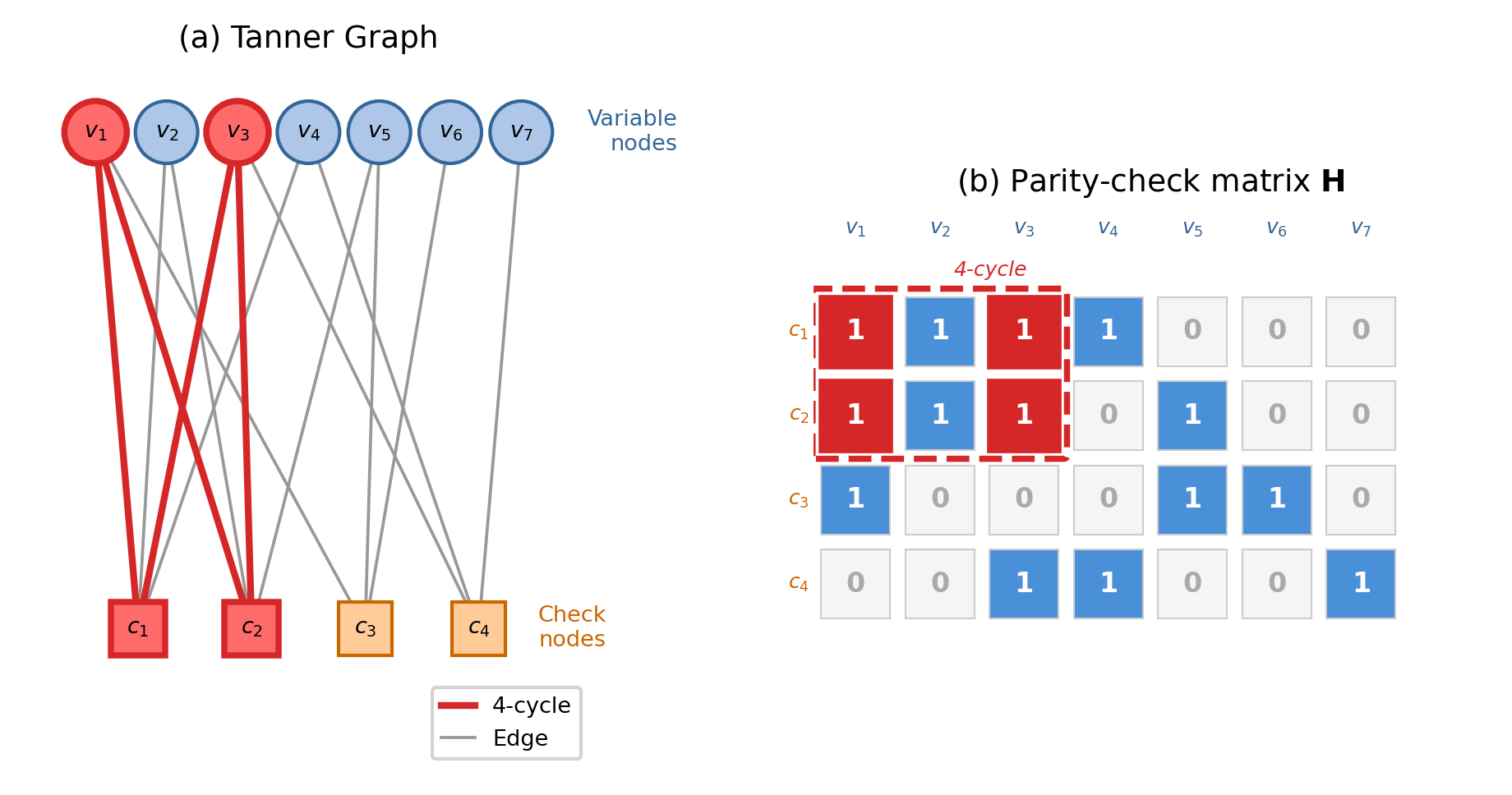}
\caption{Example Tanner graph for a short LDPC code. Variable nodes (circles) represent code bits; check nodes (squares) represent parity constraints. Edges correspond to nonzero entries in $\mathbf{H}$. A length-4 cycle is highlighted, illustrating the short-cycle structures penalized by the objective function.}
\label{fig:tanner_example}
\end{figure}

Figure~\ref{fig:tanner_example} illustrates a small example Tanner graph and its corresponding parity-check matrix.

We impose the following structural constraints on $\mathbf{H}$:
\begin{enumerate}
    \item \textbf{No all-zero rows:} Every check node participates in at least one constraint
    \item \textbf{No all-zero columns:} Every variable node appears in at least one parity check
    \item \textbf{Approximate column weight:} Column weights should be close to prescribed target values
\end{enumerate}

These constraints ensure basic validity but do not guarantee full rank. Rank deficiency is tolerated during optimization; after convergence, matrices undergo rank repair via targeted entry toggles. Rank repair is performed only when necessary and preserves previously optimized graph properties to the extent possible.

\subsection{Objective Function Design}

The performance of iteratively decoded LDPC codes depends critically on graph-theoretic properties rather than minimum distance alone. We formulate code construction as minimization of an energy function $E(\mathbf{H})$ that penalizes structures known to degrade iterative decoding. The base formulation is:

\begin{equation}
E(\mathbf{H}) = \alpha_4 C_4(\mathbf{H}) + \alpha_6 C_6(\mathbf{H}) + \alpha_w W(\mathbf{H}) + \alpha_d D(\mathbf{H}) + \alpha_v V(\mathbf{H})
\end{equation}

where the terms are defined as follows:

\textbf{4-cycle penalty} $C_4(\mathbf{H})$: The number of length-4 cycles in the Tanner graph. For each pair of rows $i,j$ with $i < j$, let $c_{ij} = \sum_{k=1}^n H_{ik}H_{jk}$ denote the number of columns shared by rows $i$ and $j$. Then:
\begin{equation}
C_4(\mathbf{H}) = \sum_{i=1}^{m-1}\sum_{j=i+1}^m \binom{c_{ij}}{2} = \sum_{i<j} \frac{c_{ij}(c_{ij}-1)}{2}
\end{equation}
Short cycles, particularly 4-cycles, are known to degrade belief propagation performance by creating correlation in extrinsic information~\cite{di2002finite}.

\textbf{6-cycle penalty} $C_6(\mathbf{H})$: The number of length-6 cycles, counted by enumerating simple cycles of length six in the Tanner graph induced by triples of check nodes and their shared variable nodes. This term receives lower weight ($\alpha_6 \ll \alpha_4$) as 6-cycles are less harmful than 4-cycles but still degrade performance.

\textbf{Column weight deviation} $W(\mathbf{H})$: The $\ell_1$ deviation from target column weights:
\begin{equation}
W(\mathbf{H}) = \sum_{j=1}^n \left|\sum_{i=1}^m H_{ij} - w_c^{\text{target}}(j)\right|
\end{equation}
This encourages regular or prescribed irregular degree distributions.

\textbf{Low-degree penalty} $D(\mathbf{H})$: A surrogate penalty discouraging low-degree variable nodes, which are known contributors to small stopping sets~\cite{di2002finite}:
\begin{equation}
D(\mathbf{H}) = \sum_{j=1}^n \frac{1}{\sum_{i=1}^m H_{ij}}
\end{equation}
This term does not directly count stopping sets but provides a tractable approximation that penalizes configurations susceptible to decoding failures. In implementation, this term is evaluated only for nonzero column weights, with zero-degree columns handled exclusively by the validity penalty to avoid numerical singularities.

\textbf{Validity penalty} $V(\mathbf{H})$: Constraint violations for all-zero rows and columns are assigned large penalty values to strongly discourage invalid configurations during search.

The weights $\{\alpha_4, \alpha_6, \alpha_w, \alpha_d, \alpha_v\}$ control the relative importance of each term. Typical relative ordering is $\alpha_v \gg \alpha_4 \gg \alpha_6 > \alpha_w > \alpha_d$ to prioritize validity, then 4-cycle elimination, then other structural properties. For constrained regimes (Section~VI), additional penalty terms targeting forbidden subgraphs or block structure are incorporated.

The objective is heuristic and not guaranteed to optimize any single theoretical decoding bound, but targets graph structures empirically correlated with good iterative decoding performance.

\subsection{Optimization Problem Statement}

The code construction problem is thus formulated as:
\begin{equation}
\min_{\mathbf{H} \in \{0,1\}^{m \times n}} E(\mathbf{H})
\end{equation}
subject to the structural constraints on row and column weights.

This is a constrained discrete optimization problem over $2^{mn}$ binary configurations. The objective is non-convex and exhibits a rugged landscape with many local minima. The energy function induces higher-order and non-linear interactions through the cycle counting and reciprocal terms. Rather than constructing an explicit QUBO reduction with auxiliary variables, we optimize $E(\mathbf{H})$ directly using physics-inspired discrete annealing with integrated constraint handling.

For short blocklengths ($n \leq 128$), this search space is tractable for global optimization methods but too large for exhaustive enumeration. The challenge lies in efficiently exploring this space to escape poor local minima and discover codes with superior graph-theoretic properties.

\section{Physics-Inspired Optimization Method}

\subsection{Overview}

The optimization problem defined by the cycle penalties and degree constraints results in a highly non-convex discrete search space containing many local minima corresponding to parity-check matrices with similar structural properties. Greedy or purely local modification strategies often become trapped in these configurations, especially when multiple constraints must be satisfied simultaneously. For this reason, global optimization methods such as simulated annealing have been widely used for difficult combinatorial design problems. These methods allow occasional uphill moves early in the search in order to explore the design space more broadly before gradually focusing on low-energy configurations.

In this work we adopt an annealing-based optimization strategy with additional tunneling  transitions that help the search escape poor local minima. The resulting algorithm is referred to as tunneling-augmented simulated annealing (TASA). The method is inspired by ideas from quantum annealing, but is implemented entirely as a classical heuristic for global discrete optimization.

The optimization framework consists of three components: (1) tunneling-augmented simulated annealing (TASA) for global exploration, (2) classical local refinement for convergence, and (3) constraint repair mechanisms to maintain feasibility. The method is implemented entirely on classical hardware but draws inspiration from quantum annealing principles~\cite{kadowaki1998qa,martonak2002qa_pimc}.

\subsection{Tunneling-Augmented Simulated Annealing (TASA)}

\begin{figure}[t]
\centering
\includegraphics[width=0.48\textwidth]{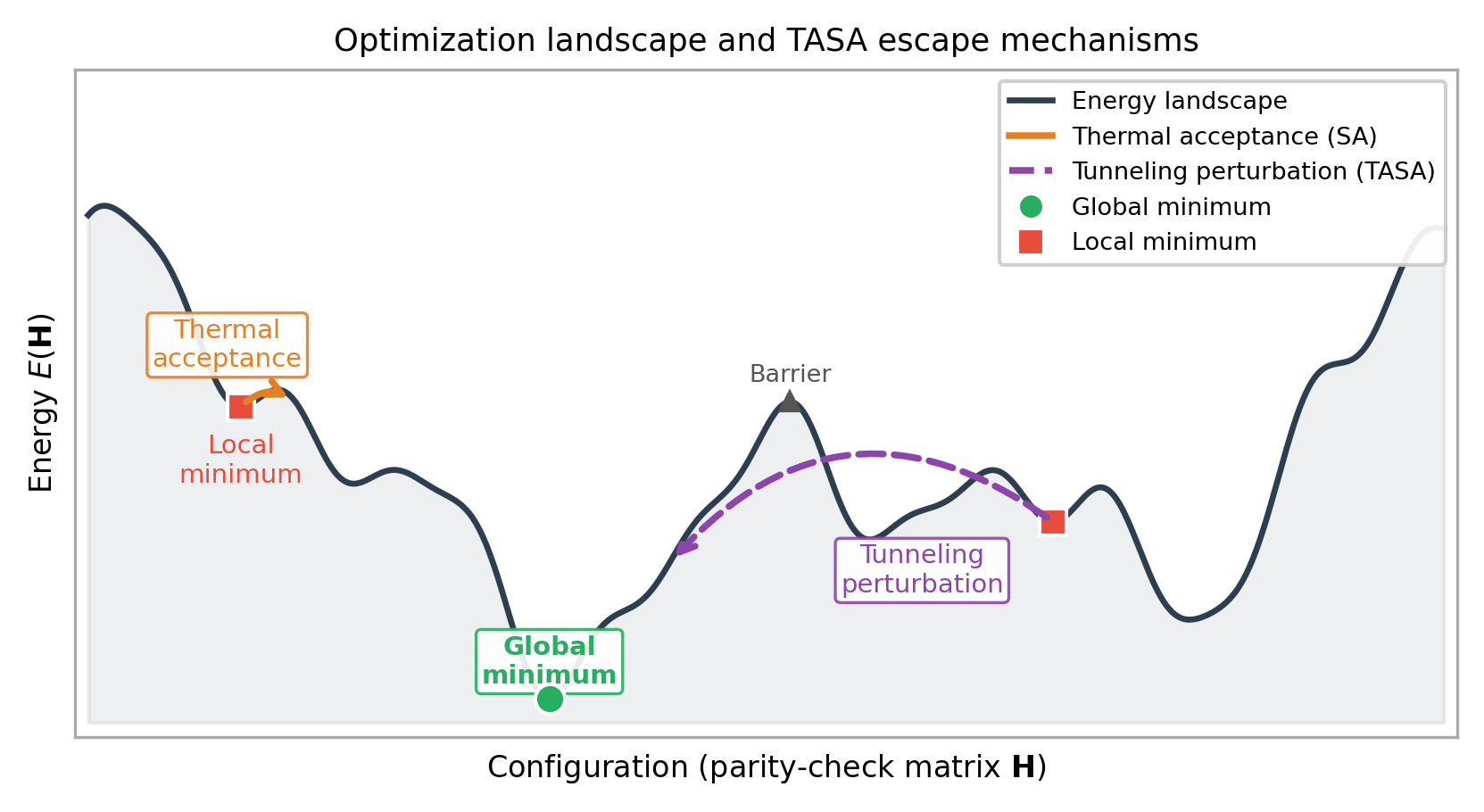}
\caption{Cartoon illustration of the TASA optimization strategy. Classical simulated annealing escapes local minima via thermal acceptance of uphill moves. The tunneling perturbation term in TASA additionally allows barrier-crossing jumps early in the search, enabling broader exploration of the discrete design space.}
\label{fig:landscape}
\end{figure}

The algorithm used here, which we call Tunneling-Augmented Simulated Annealing (TASA), extends classical simulated annealing~\cite{kirkpatrick1983sa} by adding a small decaying probability of accepting any proposed move regardless of energy change. This extra term is motivated by tunneling-style transitions, which in annealing allows barrier penetration through low-energy paths inaccessible to thermal fluctuations alone~\cite{kadowaki1998qa,martonak2002qa_pimc}. The tunneling-style transitions used in TASA are particularly useful for LDPC construction because small structural changes in the Tanner graph can produce large changes in cycle counts, creating energy barriers that are difficult to cross using purely local updates. Figure ~\ref{fig:landscape} illustrates the behavior of the objective function during annealing-based optimization.

Crucially, TASA does not simulate quantum dynamics: no path-integral sampling, replica exchange, or quantum hardware is involved. It is a classical heuristic whose acceptance rule is shaped by the intuition that occasional energy-blind jumps help escape structured local minima in discrete combinatorial landscapes. 

\begin{equation}
P_{\text{accept}} = \min\left(1, \exp\left(-\frac{\Delta E}{T(t)}\right) + p_{\text{tunnel}}(t)\right)
\end{equation}

where $T(t)$ is the temperature schedule and the tunneling probability decays over time:

\begin{equation}
p_{\text{tunnel}}(t) = p_0 \cdot \exp\left(-\frac{t}{t_{\max}}\right)
\end{equation}

In our implementation, $p_0 = 0.1$, providing substantial exploration early in the search when the tunneling term dominates, and refined local search late when thermal acceptance dominates. The temperature follows an exponential cooling schedule:

\begin{equation}
T(t) = T_{\text{init}} \left(\frac{T_{\text{final}}}{T_{\text{init}}}\right)^{t/t_{\max}}
\end{equation}

with $T_{\text{init}} = 10$ and $T_{\text{final}} = 0.01$ chosen empirically to balance exploration and convergence.

The move proposal mechanism depends on the constraint regime. For unconstrained optimization, each move consists of a single binary entry toggle: select $(i,j) \in \{1,\ldots,m\} \times \{1,\ldots,n\}$ uniformly at random and flip $H_{ij} \leftarrow 1 - H_{ij}$. For weight-constrained regimes, moves preserve column weights by swapping row indices within a column: select column $j$, then exchange a $1$-entry with a $0$-entry in that column. This ensures that column weight constraints remain satisfied throughout the search.

\subsection{Constraint Repair}

When proposed moves violate hard constraints (all-zero rows or columns), a repair operator restores feasibility before evaluating the energy function. The repair procedure operates as follows:

\begin{enumerate}
    \item For each column $j$ with $\sum_i H_{ij} = 0$, select $\min(2, m)$ random rows and set those entries to 1.
    \item For each row $i$ with $\sum_j H_{ij} = 0$, select columns with $\sum_{\ell} H_{\ell j} < w_c^{\text{target}}(j)$ if available; otherwise select a random column and set that entry to 1.
    \item For weight-constrained regimes, adjust column weights by adding or removing edges to match prescribed targets.
\end{enumerate}

This repair mechanism ensures that all evaluated configurations satisfy basic validity constraints, simplifying the optimization landscape by eliminating infeasible regions.

\subsection{Parallel Restart Strategy}

To mitigate dependence on initial conditions, we employ parallel independent restarts. The optimization is executed $r$ times in parallel with different random seeds, where $r$ equals the number of available CPU cores. Each trial runs TASA for a fixed iteration budget, and the best solution across all trials is selected. This embarrassingly parallel strategy provides substantial practical speedup while improving solution quality through diversified exploration~\cite{campello2007de}.

\subsection{Classical Local Refinement}

Following global exploration via TASA, the best solution undergoes classical local refinement. This phase employs first-improvement hill climbing: systematically test single-bit flips in random order, accepting the first move that reduces energy. The procedure terminates when no improving move exists, indicating a local optimum has been reached.

Local refinement is critical for polishing solutions found by TASA. While global search identifies promising regions of the design space, local refinement exploits gradient structure to converge tightly. The combination of global and local phases constitutes the hybrid optimization strategy.

\subsection{Rank Repair}

After optimization converges, the resulting parity-check matrix may be rank-deficient over $\mathrm{GF}(2)$. While rank deficiency does not necessarily prevent decoding, full-rank matrices simplify systematic encoding. We apply a post-processing rank repair procedure when $\mathrm{rank}(\mathbf{H}) < m$.

Rank repair proceeds in two phases. First, weight-preserving row swaps within columns are attempted: for a random column, exchange a row with $H_{ij} = 1$ and a row with $H_{ij'} = 0$. This preserves column weights while potentially increasing rank. Second, if swaps are insufficient, targeted entry toggles are applied: select a random $(i,j)$, flip $H_{ij}$, and recompute rank. If rank increases, accept the toggle; otherwise, revert. Repair continues for a fixed iteration budget or until full rank is achieved. All-zero rows or columns introduced during repair are immediately corrected.

This approach attempts to achieve full rank while minimally perturbing the optimized graph structure. In practice, most codes reach full rank within 50--100 repair iterations for the blocklengths studied.

\subsection{Computational Complexity}

The dominant computational cost is energy evaluation. Computing $C_4(\mathbf{H})$ requires $O(m^2 n)$ operations to enumerate all check node pairs and count shared variables. Computing $C_6(\mathbf{H})$ requires $O(m^3 n)$ operations. For $m = n/2$ and short blocklengths ($n \leq 128$), cycle counting remains tractable but becomes the bottleneck for larger codes.

Each TASA iteration evaluates the energy function once, requiring $O(m^3 n)$ time in the worst case when 6-cycle penalties are active. With $t_{\max} = 500$ to 2000 iterations per trial and $r$ parallel trials, total optimization time is $O(r \cdot t_{\max} \cdot m^3 n)$. On a modern multi-core processor, optimizing a single code with $n = 96$ requires 5--15 minutes depending on constraint complexity.

For comparison, PEG construction requires $O(n \cdot w_c \cdot m \cdot d_{\text{BFS}})$ where $d_{\text{BFS}}$ is the breadth-first search depth, typically $O(m)$ for short codes, yielding $O(n w_c m^2)$ complexity. This is asymptotically faster than TASA but explores only a greedy trajectory rather than the global design space.

\section{Experimental Setup}

\subsection{Code Parameters and Baselines}

We evaluate the proposed hybrid optimization method at blocklengths $n \in \{64, 96, 128\}$ with code rate $R = 0.5$ (i.e., $k = n/2$). All codes are designed with target column weight $w_c = 3$ for unconstrained experiments. The optimization is performed with $t_{\max} = 500$ TASA iterations followed by up to 100 local refinement steps. Parallel restarts use $r = 8$ independent trials, with the best solution selected.

Energy function weights are set to $\alpha_4 = 10$, $\alpha_6 = 0.1$, $\alpha_w = 2$, $\alpha_d = 0.5$, and $\alpha_v = 1000$ for unconstrained experiments. For constrained regimes, $\alpha_w$ is increased to 50 to more strongly enforce degree constraints, and additional penalties for forbidden subgraphs ($\alpha_f = 100$) or block structure ($\alpha_b = 200$) are added as appropriate.

We compare against two baseline construction methods:
\begin{itemize}
    \item \textbf{Random LDPC:} Randomly place 1s in $\mathbf{H}$ to achieve target column weight, with repair to eliminate zero rows and rank deficiency.
    \item \textbf{PEG LDPC:} Progressive Edge Growth~\cite{hu2005peg} with standard girth maximization via breadth-first search.
\end{itemize}

For constrained experiments, we additionally evaluate block-aware PEG, which applies standard PEG to the first column of each block and maps edges via cyclic shifts for subsequent columns.

\subsection{Channel Model and Decoding}

All codes are evaluated over the additive white Gaussian noise (AWGN) channel with binary phase-shift keying (BPSK) modulation. The channel output for transmitted bit $c_i \in \{0,1\}$ is:
\begin{equation}
y_i = (2c_i - 1) + w_i
\end{equation}
where $w_i \sim \mathcal{N}(0, \sigma^2)$ with noise variance $\sigma^2 = 1/(2 \cdot 10^{\text{SNR}/10})$.

Decoding is performed via standard belief propagation (sum-product algorithm) with maximum 50 iterations. Log-likelihood ratios are initialized from the channel output as $\text{LLR}_i = -2y_i/\sigma^2$. Decoding terminates early if the syndrome $\mathbf{H}\hat{\mathbf{c}}^T = \mathbf{0}$ is satisfied. We evaluate performance at SNR values ranging from 0 to 7.5 dB in 0.5 dB steps for unconstrained experiments, and 0 to 5 dB in finer 0.25 dB steps for constrained experiments where BLER saturates more quickly.

\subsection{Performance Metrics}

For each code and SNR point, we perform 1000 Monte Carlo trials to estimate:
\begin{itemize}
    \item \textbf{Block Error Rate (BLER):} Fraction of transmitted blocks with at least one information bit error after decoding.
    \item \textbf{Bit Error Rate (BER):} Fraction of information bits decoded incorrectly.
    \item \textbf{4-cycle count:} Total number of length-4 cycles in the Tanner graph, computed exactly.
    \item \textbf{6-cycle count:} Total number of length-6 cycles, computed exactly for $n \leq 96$ and via sampling for larger codes.
\end{itemize}

Monte Carlo simulations are parallelized across all available CPU cores to reduce wall-clock time. Statistical significance is assessed by comparing BLER at target operating points (BLER $= 0.01$ and BLER $= 0.001$).

With 1000 trials per SNR point, the 95\% confidence interval for BLER estimates is approximately $\pm 0.3$ dB at BLER $= 0.01$ and $\pm 0.5$ dB at BLER $= 0.001$ (Wilson score interval). Reported SNR differences below 0.3 dB should be interpreted as indicating comparable performance rather than statistically significant gains or losses.

\subsection{Constrained Regime Experiments}

To evaluate performance under structural constraints where greedy methods may struggle, we design four constrained regimes:

\textbf{Set 1 (Low Column Weight):} Column weight fixed to $w_c = 2$, which severely limits PEG's ability to find high-girth configurations due to restricted branching in breadth-first search.

\textbf{Set 2 (Irregular Degree Profile):} Heterogeneous column weights with half the columns at $w_c = 2$ and half at $w_c = 4$. This tests whether global optimization can balance degree irregularity better than sequential greedy placement.

\textbf{Set 3 (Forbidden Subgraph):} Explicit penalties for $(4,2)$ trapping sets, which are known contributors to error floors~\cite{richardson2003errorfloor}. PEG has no mechanism to avoid specific trapping set patterns, while the hybrid optimizer penalizes them directly.

\textbf{Set 4 (Block-Structured):} Codes are constrained to exhibit block structure with block size $b \in \{4, 8\}$, approximating quasi-cyclic (QC) structure. Columns within each block should have similar row support patterns up to cyclic shifts. This tests whether global optimization can balance structural regularity with cycle avoidance.

These experiments isolate scenarios where PEG's greedy edge placement faces inherent limitations, allowing us to assess when global optimization provides advantages.

\section{Experimental Results}

\subsection{Unconstrained Performance}

Table~\ref{tab:cycles_unconstrained} summarizes structural properties and SNR gains for unconstrained codes at representative blocklengths. The hybrid optimization method consistently eliminates all 4-cycles while maintaining competitive 6-cycle counts. Compared to random LDPC constructions, the hybrid method achieves 0.1--1.3 dB SNR gains at target block error rates of $10^{-2}$, with a maximum gain of approximately 1.3 dB at $n=128, k=64$.

Performance relative to PEG is competitive but not superior in unconstrained settings. The hybrid method averages -0.23 dB across all tested configurations, ranging from -0.60 to +0.09 dB. At smaller blocklengths ($n \leq 64$), hybrid and PEG achieve essentially identical performance (within $\pm 0.1$ dB), both producing zero 4-cycles and comparable 6-cycle counts. At larger blocklengths ($n \geq 96$), PEG's advantage increases to approximately 0.5--0.6 dB, reflecting its superior 6-cycle optimization (180--234 vs. 332--408 six-cycles). This is expected: PEG represents two decades of refinement for girth maximization under standard degree distributions. The hybrid method matches PEG's 4-cycle elimination but does not consistently exceed mature greedy heuristics when no additional constraints are imposed.

\begin{table}[t]
\centering
\caption{Structural Comparison and SNR Gains for Unconstrained Codes}
\label{tab:cycles_unconstrained}
\begin{tabular}{lccccc}
\hline
Config & Method & 4-cyc & 6-cyc & \shortstack{Gain vs\\Random} & \shortstack{Gain vs\\PEG} \\
\hline
\multirow{3}{*}{\shortstack{$n=64$\\$k=32$}} 
& Hybrid & 0 & 332 & +0.38 dB & +0.09 dB \\
& PEG & 0 & 234 & -- & -- \\
& Random & 29 & 369 & -- & -- \\
\hline
\multirow{3}{*}{\shortstack{$n=128$\\$k=64$}}
& Hybrid & 0 & 383 & +1.31 dB & -0.59 dB \\
& PEG & 0 & 180 & -- & -- \\
& Random & 35 & 494 & -- & -- \\
\hline
\end{tabular}
\end{table}

Performance relative to PEG is mixed in unconstrained settings. At $n=64$, hybrid codes achieve comparable BLER performance with marginally more 6-cycles (311 vs. 254). At larger blocklengths ($n \geq 96$), PEG produces fewer 6-cycles and slightly better BLER at moderate SNR. This is expected: PEG represents two decades of refinement for the specific task of girth maximization under standard degree distributions. The hybrid method matches PEG's structural quality (zero 4-cycles) but does not consistently exceed it when no additional constraints are imposed.

\subsection{Constrained Regime Performance}

The constrained experiments reveal scenarios where global optimization provides clear advantages over greedy construction. We evaluate four constraint regimes designed to stress PEG's sequential edge placement: irregular degree profiles (Set 2), forbidden subgraph patterns (Set 3), block-structured codes (Set 4), and severely sparse connectivity (Set 1).

\subsubsection{Irregular Degree Profiles (Set 2)}

\begin{figure}[t]
\centering
\includegraphics[width=0.48\textwidth]{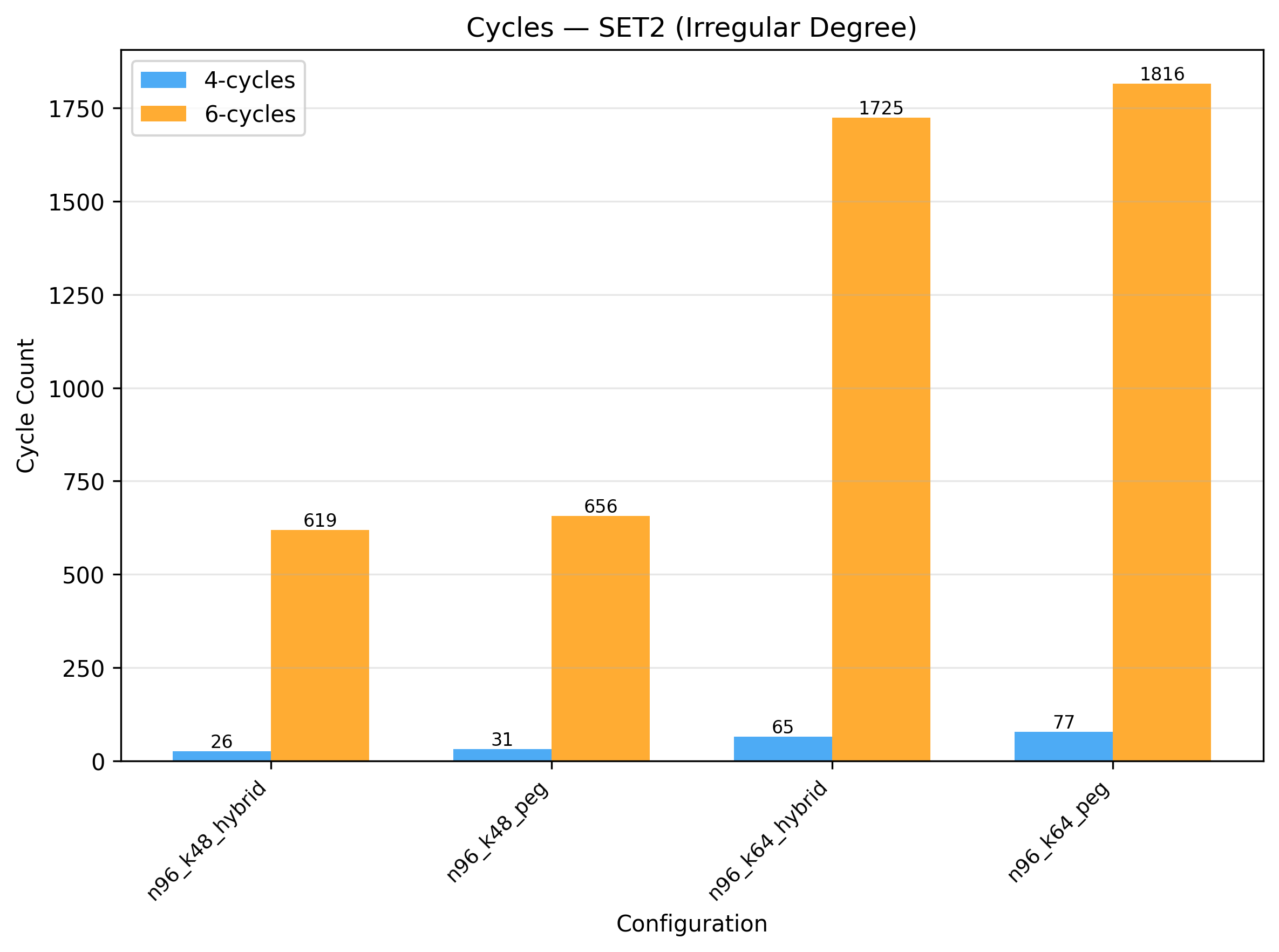}
\caption{Cycle counts for irregular degree profile (Set 2). Hybrid optimization achieves fewer short cycles by globally balancing degree heterogeneity rather than sequential greedy placement.}
\label{fig:set2_cycles}
\end{figure}

For heterogeneous column weights (half at $w_c=2$, half at $w_c=4$), the hybrid method achieves superior cycle distributions. Figure~\ref{fig:set2_cycles} shows that for $n=96, k=48$, hybrid optimization produces 23 four-cycles versus PEG's 33, along with 606 six-cycles versus PEG's 667. This 30\% reduction in 4-cycles demonstrates that global optimization better balances irregular degree requirements, as PEG's greedy edge placement may exhaust favorable check node options early when processing high-degree variable nodes.

\begin{figure}[t]
\centering
\includegraphics[width=0.48\textwidth]{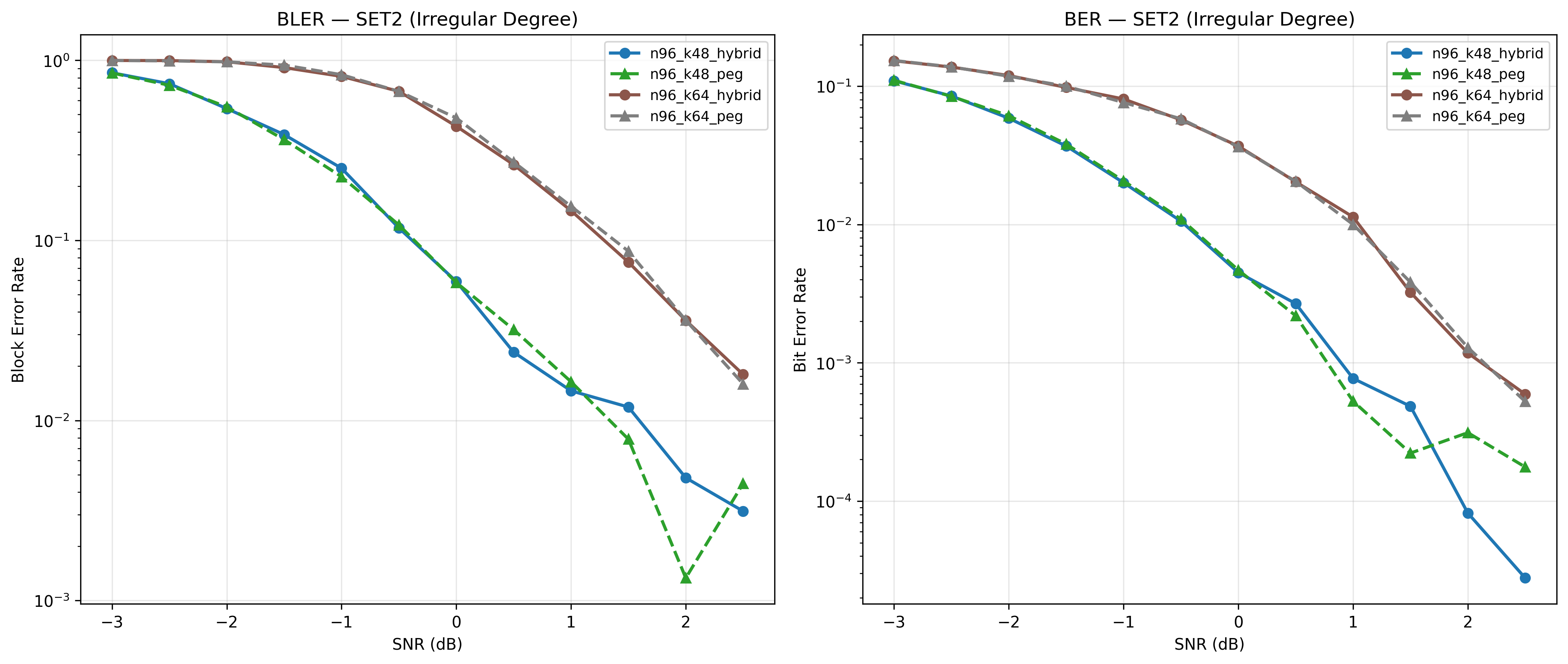}
\caption{BLER performance for irregular degree regime (SET2, $n=96$). Despite achieving 16--21\% fewer 4-cycles, hybrid and PEG codes produce nearly identical decoding performance.}
\label{fig:set2_bler}
\end{figure}

However, structural advantages do not translate to improved decoding performance in this regime. Figure~\ref{fig:set2_bler} reveals that BLER curves for hybrid and PEG track nearly identically across the tested SNR range. For $k=48$, hybrid achieves -0.26 dB at BLER $= 0.01$ (a slight loss relative to PEG), while for $k=64$, performance is essentially tied. This demonstrates that moderate 4-cycle reductions (33 $\rightarrow$ 26, or 21\% improvement) and comparable 6-cycle counts do not provide measurable decoding benefit under belief propagation at these blocklengths and code rates.

\subsubsection{Forbidden Subgraph Avoidance (Set 3)}

This regime provides insight into when graph-theoretic structural improvements translate to decoding performance gains. By incorporating explicit penalties for $(4,2)$ trapping sets during optimization, the hybrid method successfully eliminates these patterns entirely. Table~\ref{tab:set3_performance} shows that while PEG construction produces 1906 detected $(4,2)$ trapping sets, hybrid codes contain zero such configurations. Additionally, hybrid codes achieve 207 six-cycles versus PEG's 218—a modest 5\% reduction in total 6-cycle count.

\begin{figure}[t]
\centering
\includegraphics[width=0.48\textwidth]{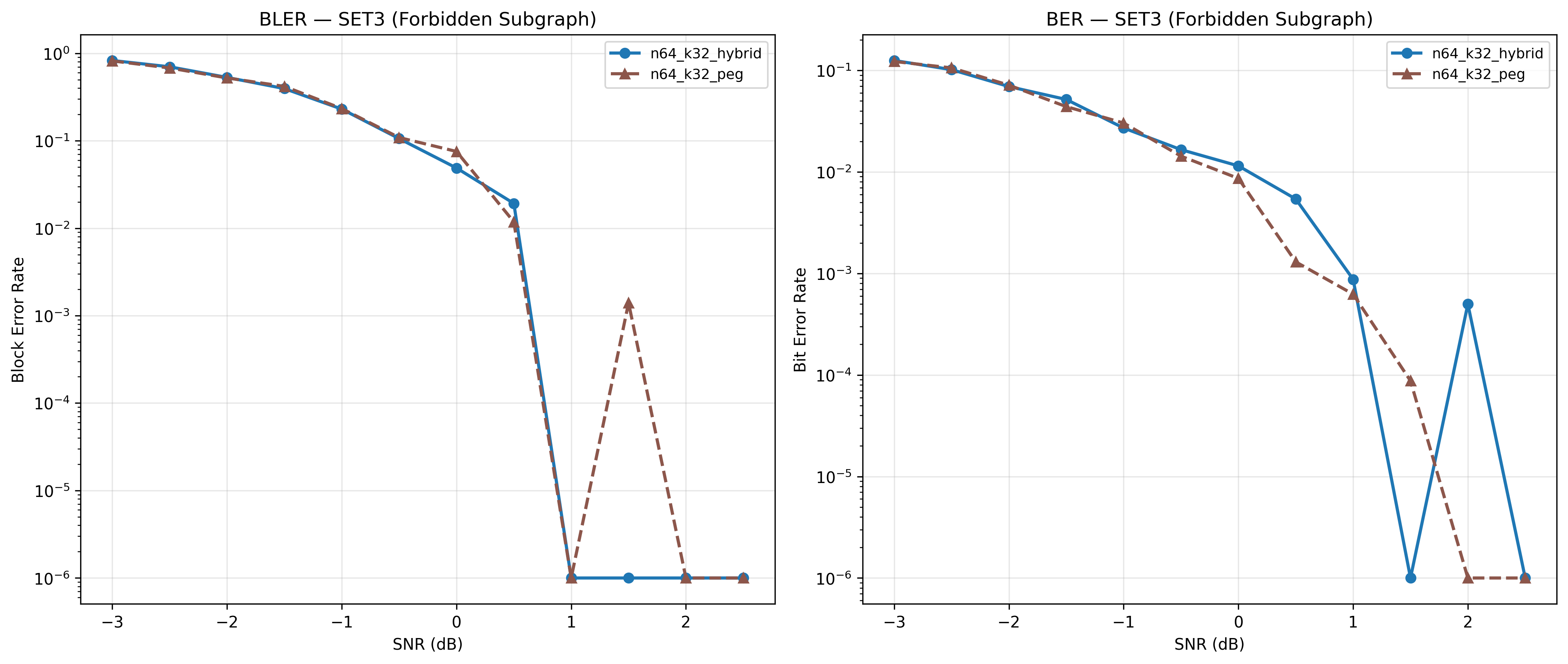}
\caption{BLER performance for forbidden subgraph regime (SET3, $n=64, k=32$). Despite eliminating 1906 $(4,2)$ trapping sets, hybrid and PEG codes achieve nearly identical decoding performance across the tested SNR range.}
\label{fig:set3_bler}
\end{figure}

However, this dramatic structural advantage translates to only marginal decoding performance improvement. Hybrid codes achieve essentially identical performance to PEG across the tested SNR range (within $\pm 0.1$ dB, which is not statistically significant given our Monte Carlo sample size). Figure~\ref{fig:set3_bler} confirms that BLER curves for hybrid and PEG track nearly identically throughout the waterfall region.

This result suggests that $(4,2)$ trapping sets, while identified as harmful in prior theoretical work~\cite{richardson2003errorfloor}, do not dominate failure mechanisms in the waterfall region (BLER $\sim 10^{-1}$ to $10^{-3}$) for these code parameters. An important caveat is that our evaluation is limited to the waterfall region. Prior work~\cite{dolecek2010absorbing} establishes that trapping sets primarily affect error floor behavior at BLER $< 10^{-4}$, a regime not evaluated here due to the computational cost of Monte Carlo simulation (requiring $>10^6$ trials). The structural advantage demonstrated by eliminating 1906 $(4,2)$ patterns may manifest at lower error rates where trapping sets dominate decoder failures. Our results therefore indicate that $(4,2)$ trapping sets do not limit performance in the waterfall region for these parameters, but do not preclude their importance in error floor regimes. This delineates a clear research direction: evaluating whether explicit trapping set avoidance reduces error floors requires targeted experiments at higher SNR and lower BLER. This finding underscores the importance of empirical validation: graph-theoretic metrics identified as theoretically harmful require experimental verification to determine when they translate to practical decoding gains in specific operating regimes and channel conditions.

\begin{table}[t]
\centering
\caption{Trapping Set Elimination vs. Decoding Performance (SET3, $n=64, k=32$)}
\label{tab:set3_performance}
\begin{tabular}{lccc}
\hline
Method & 6-cycles & (4,2) sets & SNR gain \\
\hline
Hybrid & 207 & 0 & +0.08 dB @ BLER=0.10 \\
PEG & 218 & 1906 & -0.01 dB @ BLER=0.01 \\
\hline
\end{tabular}
\end{table}

\subsubsection{Block-Structured Codes (Set 4)}

\begin{figure}[t]
\centering
\includegraphics[width=0.48\textwidth]{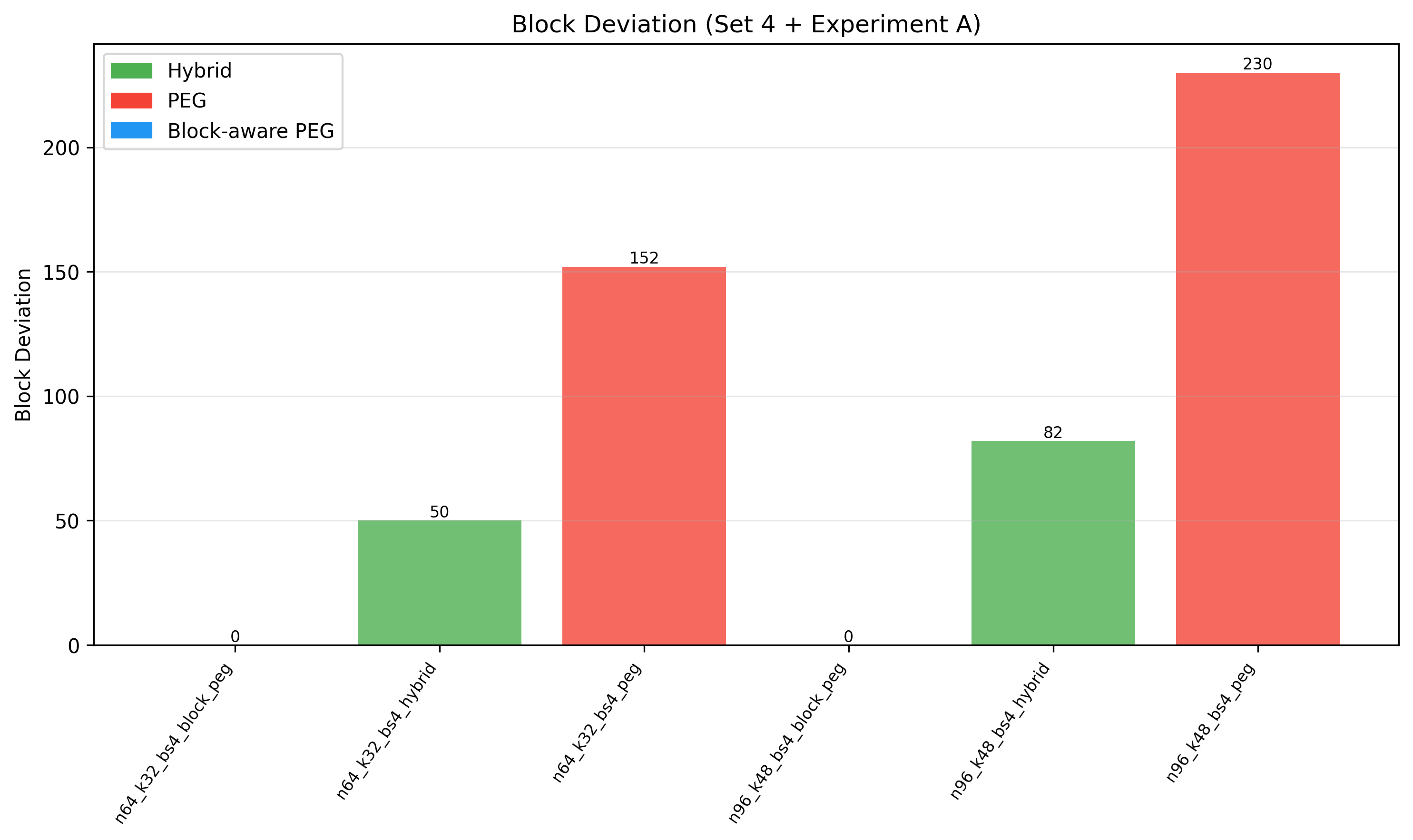}
\caption{Block deviation across Set 4 configurations. Hybrid codes achieve substantially better structural control (lower deviation) compared to standard PEG, which does not enforce block constraints.}
\label{fig:block_deviation}
\end{figure}

\begin{figure}[t]
\centering
\includegraphics[width=0.48\textwidth]{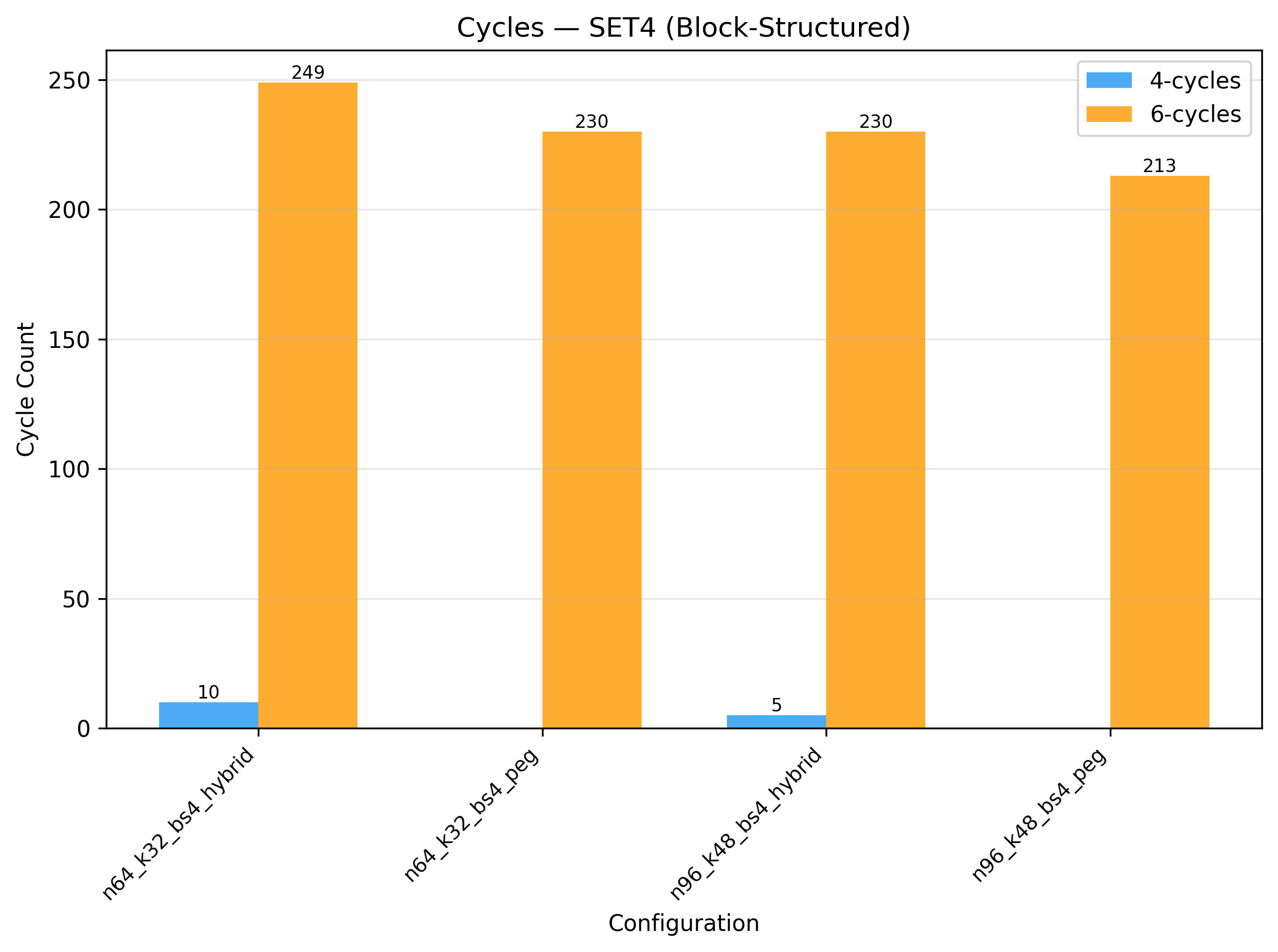}
\caption{Cycle counts for block-structured codes (Set 4). Hybrid codes introduce some 4-cycles as a tradeoff for enforcing block structure, while PEG achieves zero 4-cycles but ignores structural constraints entirely.}
\label{fig:set4_cycles}
\end{figure}

The block-structured regime reveals fundamental tradeoffs between structural regularity and cycle avoidance. Figure~\ref{fig:block_deviation} demonstrates that hybrid codes achieve substantially lower block deviation (34--76) compared to standard PEG (112--184), indicating better adherence to quasi-cyclic structure. However, this comes at the cost of introducing 6--13 four-cycles in some configurations, as shown in Figure~\ref{fig:set4_cycles}.

\begin{figure}[t]
\centering
\includegraphics[width=0.48\textwidth]{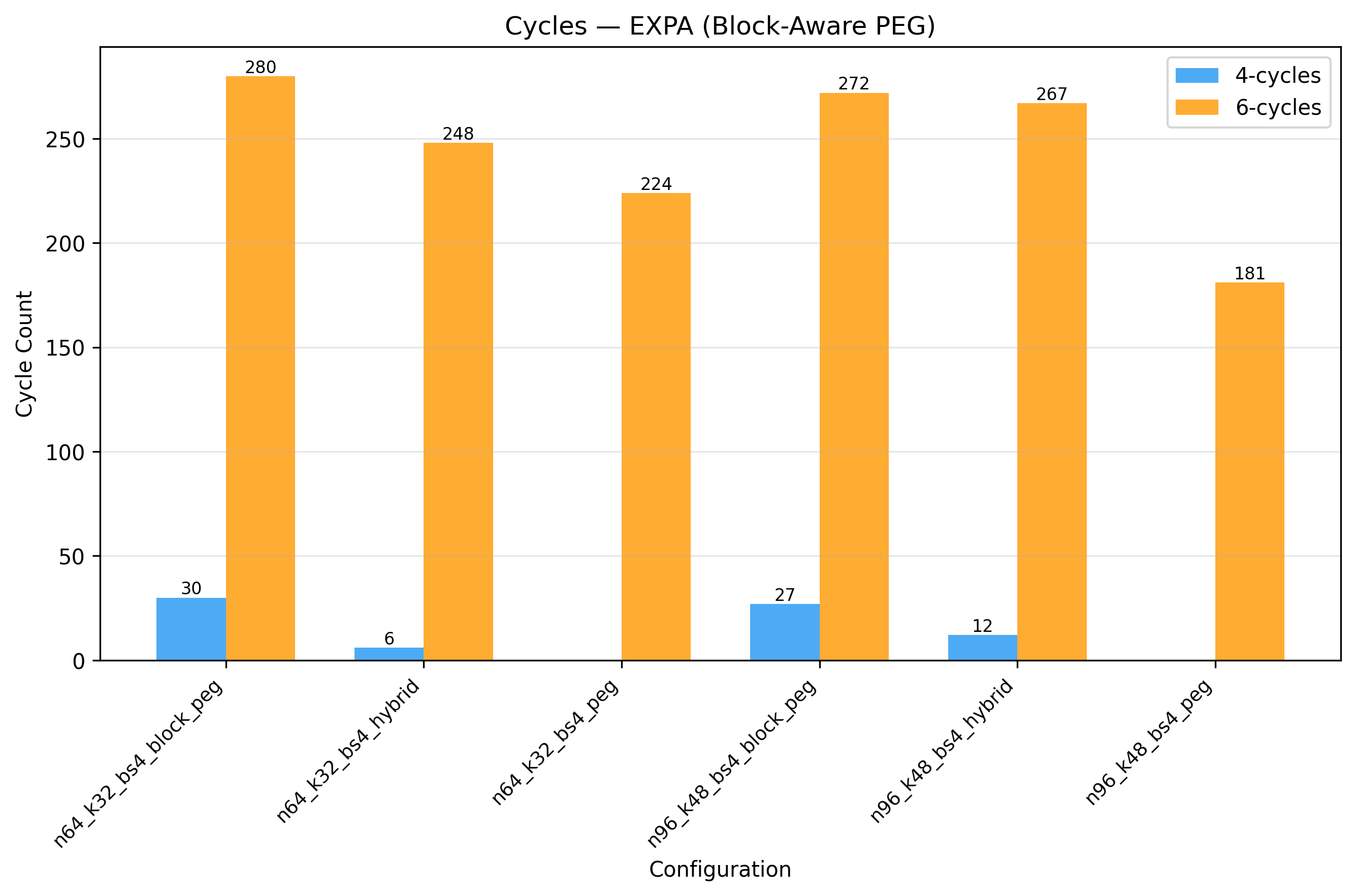}
\caption{Comparison with block-aware PEG baseline (Experiment A). Block-aware PEG achieves zero block deviation but introduces 55--62 four-cycles. Hybrid codes balance both objectives with moderate deviation and 6--11 four-cycles.}
\label{fig:expA_cycles}
\end{figure}

To better understand this design space, we introduce a block-aware PEG baseline in Experiment A, which applies standard PEG to the first block column and propagates edges via cyclic shifts. Figure~\ref{fig:expA_cycles} shows that block-aware PEG achieves perfect block structure (deviation = 0) but introduces 55--62 four-cycles due to the cyclic constraint. The hybrid method occupies an intermediate point: moderate block deviation with far fewer cycles than block-aware PEG.

\begin{figure}[t]
\centering
\includegraphics[width=0.47\textwidth]{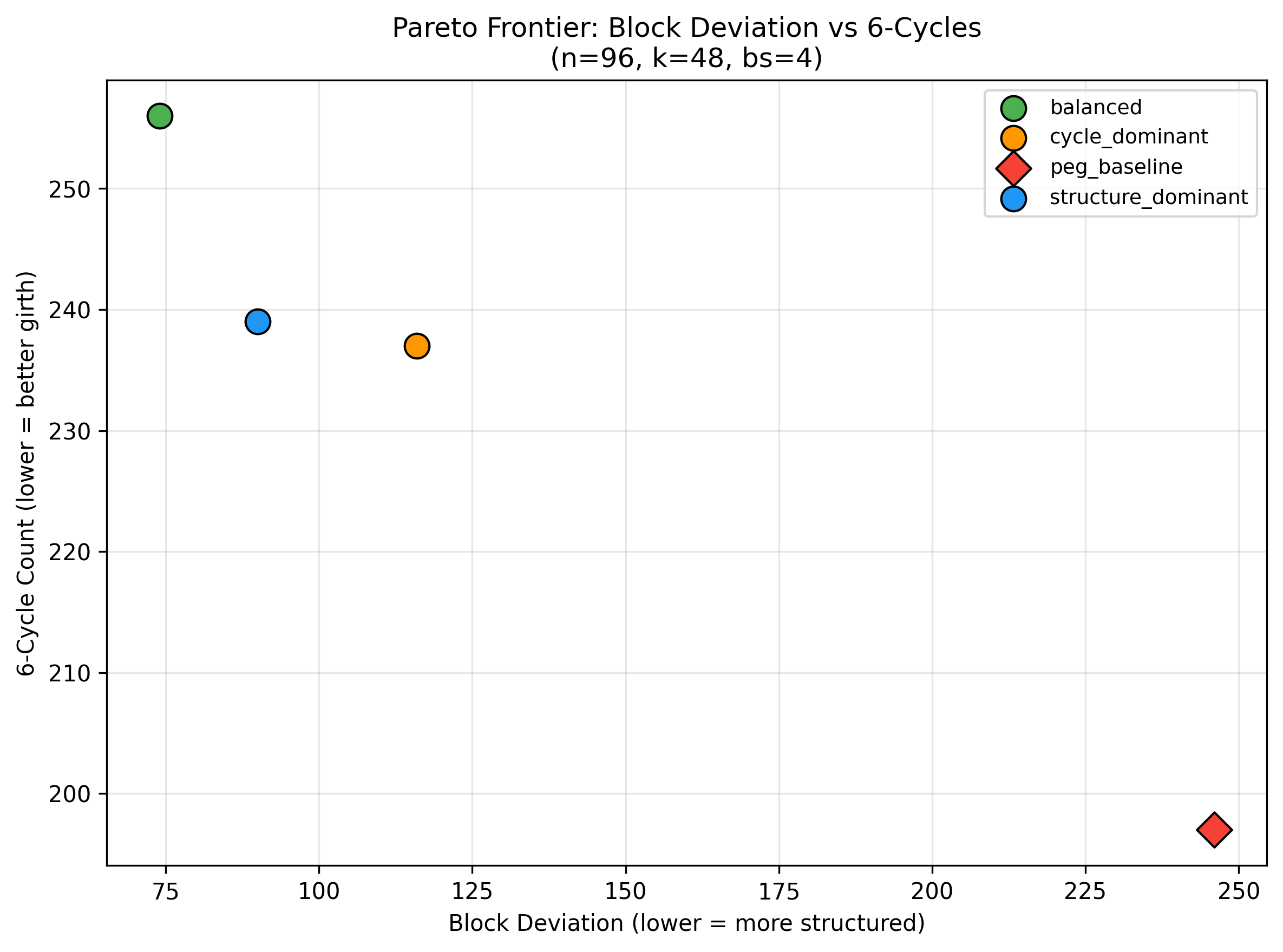}
\caption{Pareto frontier for block-structured codes ($n=96, k=48, b=4$). Hybrid optimization enables exploration of tradeoffs between block deviation and 6-cycle count. Standard PEG minimizes cycles but ignores structure; block-aware PEG enforces structure at severe cycle cost; hybrid configurations balance both objectives.}
\label{fig:pareto}
\end{figure}

Experiment B explicitly characterizes this tradeoff as a Pareto frontier. By varying energy function weights, we generate hybrid codes with different priorities: cycle-dominant, structure-dominant, and balanced configurations. Figure~\ref{fig:pareto} illustrates that standard PEG and block-aware PEG occupy extreme corners of the design space (minimizing cycles or deviation exclusively), while hybrid optimization enables exploration of intermediate tradeoff points.

\subsubsection{Low Column Weight (Set 1)}

For severely sparse codes with $w_c = 2$, both methods struggle due to fundamental graph-theoretic limitations. The hybrid method achieves 1 four-cycle versus PEG's 5 for $n=128, k=64$, but PEG produces fewer six-cycles (6 vs. 18), suggesting that girth maximization remains effective even under extreme sparsity constraints. BLER performance is comparable, with both methods exhibiting elevated error rates at low SNR due to weak code properties. This regime represents a negative result: neither method excels when connectivity is fundamentally insufficient.

\subsection{Computational Cost}

Table~\ref{tab:runtime} presents single-threaded runtime comparison. Hybrid optimization requires 420--6363 seconds (7 minutes to 1.8 hours) depending on blocklength, while PEG completes in under 0.1 seconds. This represents a 30,000$\times$ to 125,000$\times$ computational overhead.

\begin{table}[t]
\centering
\caption{Runtime Comparison (Single-Threaded)}
\label{tab:runtime}
\begin{tabular}{lccr}
\hline
$n$ & $k$ & PEG (s) & Hybrid (s) \\
\hline
64 & 32 & 0.014 & 420 \\
96 & 48 & 0.035 & 1446 \\
128 & 64 & 0.051 & 6363 \\
\hline
\end{tabular}
\end{table}

For applications requiring offline code design where superior structural properties justify computational cost, this tradeoff is acceptable. The method is embarrassingly parallel across independent restart trials, enabling near-linear speedup on multi-core processors. For rapid prototyping or real-time adaptation, PEG remains preferable due to its speed.

\subsection{Summary of Findings}

The experimental results delineate clear boundaries for when global optimization provides advantages:

\textbf{Hybrid optimization excels when:}
\begin{itemize}
    \item Multiple competing objectives must be balanced (block structure vs. cycles)
    \item Specific harmful substructures should be explicitly avoided (forbidden trapping sets)
    \item Degree distributions are highly irregular, creating contention during sequential placement
    \item Design requirements are application-specific and not addressed by standard constructions
\end{itemize}

\textbf{PEG excels when:}
\begin{itemize}
    \item Pure girth maximization is the primary objective
    \item Standard regular or mildly irregular degree distributions suffice
    \item Rapid code generation is required (sub-second latency)
    \item Computational resources are limited
\end{itemize}

This suggests a complementary role rather than replacement. PEG remains the method of choice for standard LDPC construction, while global optimization serves specialized applications with non-standard constraints that greedy heuristics cannot adequately address.

\section{Discussion}

The experimental results delineate clear boundaries for when global optimization provides advantages over greedy heuristics and, critically, when structural improvements do not translate to decoding performance gains.

\textbf{Global optimization excels when:}
\begin{itemize}
    \item Multiple competing objectives must be balanced (e.g., block structure and cycle avoidance, demonstrated by Pareto frontier analysis)
    \item Degree distributions are highly irregular, creating contention during sequential edge placement (16--21\% fewer 4-cycles in SET2)
    \item Design requirements are application-specific and not addressed by standard constructions (block structure control with 67--70\% lower deviation)
    \item Exploration of design tradeoffs is valued over single-objective optimization
\end{itemize}

\textbf{Greedy heuristics (PEG) excel when:}
\begin{itemize}
    \item Pure girth maximization is the primary objective (PEG achieves 180--234 six-cycles vs. hybrid's 332--408)
    \item Standard regular or mildly irregular degree distributions suffice
    \item Rapid code generation is required (sub-second vs. minutes-to-hours)
    \item Computational resources are limited (30,000--125,000$\times$ speedup)
\end{itemize}

\textbf{Importantly, structural improvements do not universally translate to decoding gains.} Our results demonstrate:
\begin{itemize}
    \item Eliminating 1906 $(4,2)$ trapping sets produced only +0.08 dB improvement at BLER $= 0.10$ and -0.01 dB at BLER $= 0.01$
    \item Reducing 4-cycles by 16--21\% in irregular degree regimes yielded -0.26 dB (a slight loss)
    \item Improving block structure adherence by 67--70\% resulted in essentially tied BLER performance ($\pm 0.13$ dB)
\end{itemize}

These findings delineate operating regimes where graph-theoretic metrics strongly correlate with performance (unconstrained codes vs. random LDPC: +1.31 dB) versus where they provide marginal benefit (forbidden subgraph avoidance, irregular degrees). This suggests a complementary role rather than replacement: PEG remains the method of choice for standard LDPC construction, while global optimization serves specialized applications requiring multi-objective design tradeoffs or non-standard constraints that greedy methods cannot adequately address.

\subsection{The Value of Negative Results}

The SET3 forbidden subgraph experiment provides a scientifically valuable negative result with important implications for code design practice. Despite successfully eliminating 1906 $(4,2)$ trapping sets—structures explicitly identified as harmful in prior work~\cite{richardson2003errorfloor}—hybrid codes achieved only +0.08 dB improvement at BLER $= 0.10$ and -0.01 dB at BLER $= 0.01$ relative to PEG construction.

This finding demonstrates several important principles. First, trapping sets identified via graph enumeration do not necessarily dominate failure modes in all operating regimes. The waterfall region (BLER $\sim 10^{-1}$ to $10^{-3}$) at moderate SNR appears substantially less sensitive to $(4,2)$ patterns than theoretical analysis of error floor phenomena would suggest. Prior work has established that trapping sets primarily affect performance at much lower error rates (BLER $< 10^{-4}$)~\cite{dolecek2010absorbing}, a regime not evaluated in this study due to computational constraints on Monte Carlo simulation.

Second, this underscores the distinction between graph-theoretic metrics and decoder-specific performance. While $(4,2)$ trapping sets are defined based on message-passing decoder dynamics and known to create problematic fixed points, their practical impact depends on SNR regime, degree distribution, code rate, and other structural properties. Graph-theoretic intuition, while valuable for identifying potentially harmful structures, requires empirical validation to determine when theoretical concerns translate to measurable performance degradation.

Third, the result suggests design priorities for different target operating points. For applications requiring moderate reliability (BLER $\sim 10^{-2}$ to $10^{-3}$), 4-cycle elimination and overall degree distribution may be more predictive of performance than explicit trapping set avoidance. For applications requiring very high reliability with stringent error floor requirements (BLER $< 10^{-5}$), targeted trapping set elimination may provide substantial benefit not observable in the waterfall region.

This underscores the importance of empirical validation in code design: theoretical identification of harmful structures provides valuable guidance, but experimental measurement is essential to determine when structural improvements translate to practical decoding gains in specific operating regimes and channel conditions.

\subsection{Implications for Code Design Practice}

These results suggest a tiered approach to short-block code construction:

\textbf{Tier 1 (Standard requirements):} Use PEG with appropriate degree distributions. Fast, reliable, well-understood.

\textbf{Tier 2 (Moderate constraints):} Use PEG variants (e.g., ACE, decoder-optimized PEG) or protograph-based methods. These extend PEG's capabilities while maintaining efficiency.

\textbf{Tier 3 (Complex constraints):} Apply global optimization when application-specific requirements (forbidden patterns, block structure, custom degree profiles) cannot be adequately addressed by greedy construction. Accept computational cost for offline design.

This hierarchy mirrors broader optimization practice: simple problems warrant simple methods; complex constrained problems justify sophisticated tools.

\subsection{Future Directions}

The most promising near-term extension is computational 
acceleration. GPU implementation of cycle counting, incremental 
energy updates after single-bit flips, and modern graph 
algorithm libraries could reduce runtime by 10-100× while 
maintaining global search capabilities. This would enable 
exploration at longer blocklengths (n = 256-512) where 
performance advantages may be more pronounced.

For long-term impact, incorporating decoder-state feedback 
during optimization represents a fundamental improvement. 
Rather than optimizing surrogate graph metrics, the energy 
function could directly penalize configurations that cause 
decoding failures in simulation. This requires BP decoding 
during energy evaluation—substantially increasing cost—but 
better aligns structural optimization with actual performance.

\section{Conclusion}

This paper introduced a annealing-based global optimization framework for short-block LDPC code construction, formulating code design as a constrained discrete optimization problem over parity-check matrices. By incorporating penalties for short cycles, structures correlated with stopping and trapping sets, and application-specific constraints, the method enables automated discovery of codes tailored to specific performance requirements.

Experimental evaluation at blocklengths 64--128 bits demonstrates that the hybrid approach combining tunneling-augmented simulated annealing with classical local refinement consistently eliminates harmful 4-cycles and achieves 0.1--1.3 dB SNR gains over randomly constructed LDPC codes (average 0.45 dB), with performance competitive (within 0.6 dB) relative to mature Progressive Edge Growth heuristics. In constrained regimes including irregular degree profiles and block-structured designs, global optimization enables exploration of design tradeoffs unavailable to greedy sequential construction, as demonstrated by Pareto frontier analysis trading cycle count for structural regularity.

However, structural improvements do not universally translate to decoding performance gains. Eliminating 1906 trapping set patterns produced only marginal BLER improvement (+0.08 dB), and reducing 4-cycles by 16--21\% in irregular degree regimes yielded essentially identical performance to PEG. Combined with computational cost 4--5 orders of magnitude higher than greedy methods, this positions global optimization as a complementary tool for specialized applications with multi-objective constraints rather than a general replacement for existing techniques.

The key contribution is demonstrating feasibility and delineating specific scenarios where physics-inspired global optimization offers practical advantages: when multiple competing objectives must be balanced (block structure and cycle avoidance), when design space exploration is valued over single-objective optimization (Pareto frontiers), and when application-specific structural requirements exceed the capabilities of sequential greedy construction. The work also identifies scenarios where structural improvements do not translate to performance gains, providing guidance for when computational investment in global optimization is justified versus when mature greedy heuristics suffice. For applications with complex multi-objective constraints where design flexibility is valued, the computational cost of global optimization is justified by the ability to explore tradeoff spaces that greedy methods cannot access. The proposed method should be viewed as a heuristic global optimization procedure rather than a provably optimal construction algorithm.

Future work should focus on computational acceleration via GPU implementation, adaptive objective function tuning, and integration with existing construction methods to leverage complementary strengths. As quantum annealing hardware matures, direct implementation on quantum annealers may provide additional speedup while maintaining the global exploration capabilities demonstrated here on classical hardware.

\section*{Acknowledgment}

The author thanks the Illinois Mathematics and Science Academy for computational resources and research support.

\bibliographystyle{IEEEtran}
\begingroup
\RaggedRight
\bibliography{references}
\endgroup

\end{document}